\documentclass[12pt]{article}

\pdfoutput=1

\usepackage{graphics}
\usepackage{amssymb}
\usepackage{amsmath}
\usepackage{hyperref}
\usepackage{bbold}
\usepackage{tikz}
\usepackage{cite}

\usepackage{todonotes}

\newcommand{\beq}{\begin{equation}}

\newcommand{\eeq}{\end{equation}}
\newcommand{\bea}{\begin{eqnarray}}
\newcommand{\eea}{\end{eqnarray}}

\begin{document}

\begin{center}
${}$\\
\vspace{100pt}
{ \Large \bf Curvature Correlators in Nonperturbative
\\ \vspace{10pt} 2D Lorentzian Quantum Gravity 
}

\vspace{36pt}

{\sl J.\ van der Duin}$\,^{\dagger}$ and {\sl R.\ Loll}$\,^{\dagger,\star}$

\vspace{18pt}
{\footnotesize

$^\dagger$~Institute for Mathematics, Astrophysics and Particle Physics, Radboud University \\ 
Heyendaalseweg 135, 6525 AJ Nijmegen, The Netherlands.\\ 

\vspace{5pt}
{\it and}\\
\vspace{5pt}

$^\star$~Perimeter Institute for Theoretical Physics,\\
31 Caroline St N, Waterloo, Ontario N2L 2Y5, Canada.\\
}
\vspace{24pt}

\end{center}


\begin{center}
{\bf Abstract}
\end{center}

\noindent 
Correlation functions are ubiquitous tools in quantum field theory from both a fundamental and a practical point of view.
However, up to now their use in theories of quantum gravity beyond perturbative and asymptotically flat regimes has been limited, 
due to difficulties associated with diffeomorphism invariance and the dynamical nature of geometry. We present an
analysis of a manifestly diffeomorphism-invariant, nonperturbative two-point curvature correlator in two-dimensional Lorentzian 
quantum gravity. It is based on the recently introduced quantum Ricci curvature and uses
a lattice regularization of the full path integral in terms of causal dynamical triangulations. We discuss some of the subtleties and ambiguities
in defining connected correlators in theories of dynamical geometry, and provide strong evidence from Monte Carlo simulations
that the connected two-point curvature correlator in 2D Lorentzian quantum gravity vanishes. This work paves the way for an analogous 
investigation in higher dimensions.

\vspace{12pt}
\noindent

\newpage

\section{Introduction}
\label{sec:intro}

With tools at our disposal to evaluate the gravitational path integral fully nonperturbatively in four dimensions, 
we can take an approach to quantum gravity that focuses on concrete physical questions rather than formal structural issues. 
The tried and tested method we have in mind here is lattice gravity based on causal dynamical triangulations (CDT), 
where the path integral is obtained as the scaling limit of a regularized sum over geometries \cite{review1,review2,ency}.
CDT successfully navigates the sub\-tle\-ties of applying nonperturbative lattice methods to quantum gravity, including Wick rotation, conformal divergence 
and diffeomorphism invariance, and relies only on standard concepts and principles of quantum field theory.
An invaluable part of CDT's tool box are powerful Markov chain Monte Carlo methods, which can be and have been used extensively to compute
gauge-invariant\footnote{We will use the notions of diffeomorphism invariance and gauge invariance interchangeably.} observables in a near-Planckian regime. 

Our best bet for relating quantum gravity at the Planck scale to the real world is arguably in early-universe cosmology. 
Here, CDT lattice gravity offers a promising perspective since it has produced strong evidence that the 
dimensionality \cite{Ambjorn2004,Ambjorn2005,Ambjorn2005a}, overall shape \cite{Ambjorn2007,Ambjorn2008,Ambjorn2011a} 
and average curvature \cite{qrc3} of the quantum geometry 
emerging from the path integral are all compatible with those of a classical de Sitter universe. 
The presence of such quasi-classical behaviour (in the sense of expectation values) is quite remarkable,
because the path integral is nonperturbative and manifestly background-independent, and local quantum fluctuations of geometry 
are large in the regime that is accessible to simulations. 

An intrinsic difficulty in relating the nonperturbative quantum theory to cosmological considerations are the different scales involved and the distinct theoretical
set-ups, including their associated observables. The early universe is usually described in terms of a smooth, de Sitter-like background metric with perturbative 
quantum fluctuations. Diffeomorphism-invariant observables can be constructed in a variety of ways, using gauge-fixing, gravitational dressing and relational 
constructions (see e.g.\ \cite{donnelly,propqft} and references therein).
Gauge-invariant ob\-ser\-va\-bles exist also in a nonperturbative regime, but must be constructed differently
due to the lack of a smooth manifold structure and the presence of large quantum fluctuations.
The nature of the quantum geometry observed in CDT so far, for universes with a linear diameter $\lesssim 20\,\ell_\mathrm{Pl}$, 
suggests that at these scales a local tensorial description like in general relativity is neither adequate nor feasible \textit{in principle}. 

The most straightforward obser\-va\-bles on the ensemble of nonsmooth metric spaces of the CDT path integral  -- nowhere differentiable spaces obtained in the continuum
limit of piecewise flat triangulations -- are given by spacetime averages of (quasi-)local properties. 
Concrete examples are the above-mentioned observables whose expectation values 
match a classical limit, namely averages of the local Hausdorff and spectral
dimensions, the volume and the so-called quantum Ricci curvature \cite{qrc1,qrc2}. 

Apart from providing valuable evidence for the existence of a classical limit, these observables also capture genuine quantum signatures,
as is illustrated by the spectral dimension, whose expectation value at short distances exhibits a characteristic ``dynamical dimensional reduction" away
from its classical, large-scale value of 4 to a value compatible with 2 near the Planck scale \cite{spectral}. This purely nonperturbative phenomenon, 
discovered in CDT, has since been
corroborated qualitatively in other formulations and conjectured to be a universal property of quantum gravity \cite{carlip}.
However, there are at this stage no compelling ideas on how to relate this result to phenomenology at lower energies.

Let us emphasize that the availability of observables in nonperturbative quantum gravity \textit{and} the ability to make quantitative
statements about them amounts to significant progress
in a field long characterized by an absence of computational tools beyond perturbation theory and by a reliance on abstract principles and preconceptions 
on what a fundamental theory of quantum gravity should (or should not) look like \cite{oriti,foundations,armas,Loll2022}.
To strengthen this line of exploration further we would like to construct additional observables, which (i) are within the computational reach of current lattice methods,
(ii) provide a more fine-grained understanding of quan\-tum geo\-metry, beyond global averages, and (iii) allow for a more direct connection with 
quantities relevant in the description of the early universe.
 
The subject of this paper is an observable that meets these requirements, namely a \textit{diffeomorphism-invariant curvature correlator}.
More precisely, we will investigate a prototype in two spacetime dimensions of the connected two-point correlator of the quantum Ricci scalar curvature 
in the nonperturbative gravitational path integral defined in terms of CDT. Its novelty is two-fold: correlation functions of nontrivial geometric observables
have not been studied previously in a Lorentzian, causal lattice formulation, 
and until recently no well-defined notion of curvature was available in this context.  
The problem with the standard deficit-angle curvature inherited from Regge calculus is its divergent character in
four-dimensional (C)DT (see \cite{curvsumm} and references therein). By contrast, the quantum Ricci curvature, first introduced in \cite{qrc1}, 
has been shown to be applicable in a nonsmooth, Planckian regime in four dimensions \cite{qrc3}, while on smooth manifolds it allows one to the recover the standard
Ricci curvature \cite{qrc2}.\footnote{The quantum Ricci curvature is designed for positive definite metric spaces, including the
piecewise flat simplicial spaces of the CDT path integral obtained after the Wick rotation.}

The aim of the present work is to implement and measure the correlator of the new quantum Ricci curvature for the first time, 
as a nontrivial test case and a stepping stone toward the physically relevant case in four dimensions.
As we will see below, even in two dimensions the construction and analysis of this observable presents interesting conceptual issues
and technical challenges, providing a benchmark for what will be required in higher dimensions. 

The remainder of this paper is organized as follows.
In Sec.\ \ref{sec:found} we discuss some generalities on correlation functions in 
quantum gravity, with reference to previous, related work in nonperturbative lattice formulations. 
Sec.\ \ref{sec:setup} recalls the main ingredients of formulating 2D Lorentzian quantum gravity as the continuum limit of a lattice regularization
in terms of CDT, including the elementary building blocks, gluing rules, regularized path integral, Wick rotation, Monte Carlo moves and
how to compute observables. In Sec.\ \ref{sec:qrc} we summarize important properties and applications of the quantum Ricci curvature, both in the continuum and
on the lattice. In Sec.\ \ref{sec:correl} 
we construct geometric two-point correlators and present our Monte Carlo measurements of curvature correlators in 2D CDT quantum gravity. 
We discuss different choices for their normalization and the subtraction procedure to obtain corresponding connected correlators, which are present
due to the dynamical nature of geometry.
We conclude in Sec.\ \ref{sec:future} with a summary and outlook. Two appendices contain technical details on how to determine the onset of finite-size
effects and how to uniformly sample point pairs when measuring two-point correlators.

\section{Correlation functions in quantum gravity}
\label{sec:found}

The physical content of a relativistic quantum field theory is captured entirely by the set of its \textit{n-point correlation functions} 
or \textit{correlators}, the vacuum expectation
values of $n$-fold products of its local field operators. This is not only true for quantum fields on Minkowski spacetime, but also in a Wick-rotated 
formulation on flat Euclidean $\mathbb{R}^4$, which is related to its Lorentzian counterpart by the Osterwalder-Schrader reconstruction theorem \cite{mm}. 
Also in the context of gravitational theories, $n$-point functions play an important role. Examples are correlators in perturbative quantum gravity \cite{shapiro,christiansen} and correlation functions of quantum fields on a cosmological background \cite{qftcs,propqft}. 

What may be less well known is that correlation functions can also be defined in nonperturbative quantum gravity, 
in a manifestly background-independent and diffeomorphism-invariant way \cite{higherd,4dEDT}.
Consider the case $n\! =\! 2$, which we will focus on in this work. 
Given a local geometric scalar quantity ${\cal O}[g]$, say, a curvature scalar depen\-ding on the metric $g_{\mu\nu}$, the formal continuum expression for the
(unnormalized) diffeo\-morphism-invariant two-point correlator of $\cal O$ is given by

\begin{equation}
\langle G[{\cal O},{\cal O}](r)\rangle\! =\! \int \! {\cal D}[g]\, {\rm e}^{\, i S[g]}\! \int_M\!\! d^4x \sqrt{| g(x)|} \int_M\!\! d^4y \sqrt{| g(y)|}\, 
{\cal O}(x){\cal O}(y)\, \delta (d_g(x,y)\! -\! r),
\label{2point}
\end{equation} 
\noindent where the functional integration is over the space of all geometries $[g]$ (metrics $g$ modulo diffeomorphisms on the given 
manifold $M$), with weights depending on the gravitational action $S[g]$, and $|g(x)|$ denotes the absolute value of the determinant of the metric 
$g$ at the point $x$. Diffeomorphism invariance requires
that the distance $r$ appearing inside the Dirac $\delta$-function 
is given in terms of the invariant geodesic distance $d_g$ associated with the geometry $g$, and that the points $x$ and $y$ are integrated 
over spacetime, subject to the constraint of being a distance $r$ apart. 
The viability of such diffeomorphism-invariant two-point functions, as well as an analogous version for matter correlators on fluctuating geometries, 
has been demonstrated in models of pure and matter-coupled Euclidean quantum gravity in two dimensions, formulated in terms of
Dynamical Triangulations (DT) \cite{Ambjorn1995,Ambjorn1995a,Aoki,qmatter,Ambjorn1997a,conncorr}. 

Geometric two-point functions have also been investigated in four dimensions, again in the context of Euclidean quantum gravity based on 
DT, with inconclusive results \cite{bakkersmit,focusfp} (see also \cite{bassler}). Apart from the fact that it is unclear how to relate the underlying
Euclidean to the
physical, Lorentzian path integral, these pioneering studies suffer from at least two serious shortcomings, the absence to date of 
second-order phase transitions in 4D Euclidean quantum gravity, a necessary prerequisite for the existence of a continuum theory, and
the absence of a well-defined notion of local curvature, to investigate invariant curvature-curvature correlators. The scalar curvature operator $\cal O$ used
in previous studies of 4D Euclidean quantum gravity was based on the deficit-angle prescription for curvature due to Regge \cite{regge}. 
However, as already mentioned in Sec.\ \ref{sec:intro}, this notion of curvature is not well-defined in the nonperturbative quantum theory, because
it diverges in an uncontrolled manner in the infinite-volume limit \cite{4deu,curvsumm}. It is therefore unphysical and
not suited for computing curvature correlators. 

In this work, we exploit the progress that has been made since these early investigations. On the one hand, we will use the Lorentzian path integral
as a starting point, whose lattice formulation in terms of CDT in four dimensions
possesses several second-order phase transition lines \cite{transition,transition1,newphasecgj}. On the other hand, we will
use the novel quantum Ricci curvature, which has been shown to have an improved short-distance (UV) behaviour \cite{qrc3}. 
However, it is significantly more involved computationally,
which means that its implementation requires particular attention.

\section{Theoretical and computational set-up}
\label{sec:setup}

\begin{figure}[t]
\centering
\includegraphics[width=0.9\textwidth]{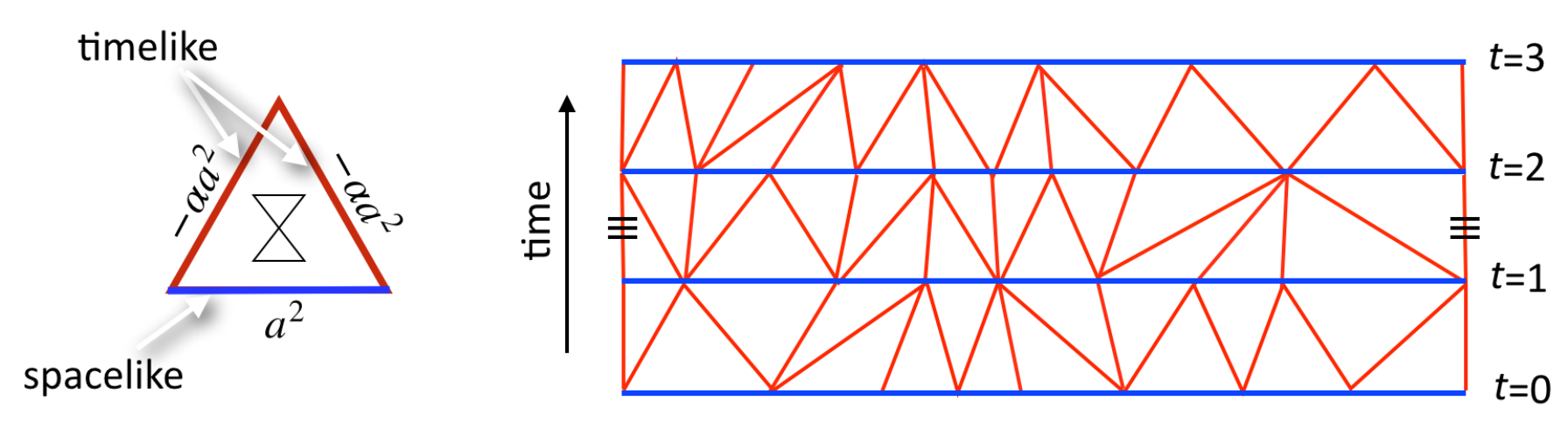}
\caption{Minkowskian building block of 2D CDT, including a lightcone (left), and a piece of a CDT spacetime, consisting of a sequence of three triangulated strips, labelled by
discrete proper time $t$ (right). Left and right boundary should be identified as indicated, leading to a cylindrical topology $I\times S^1$.}
\label{fig:tristrip}
\end{figure}

The setting for our investigation of curvature correlators is the two-dimensional Lorentzian gravitational path integral, regularized in terms of CDT, which was first
defined and solved analytically in \cite{Ambjorn1998}. It serves as a dimensionally reduced and simplified toy model of full quantum gravity. 
We will summarize its essential properties below, and refer to \cite{Ambjorn2001,review1} for further technical details. 
Recall that classical gravity in two dimensions is trivial in the sense that for fixed topology -- which we will assume -- the Einstein-Hilbert action reduces 
to a volume term proportional to a cosmological constant $\Lambda$, and there are no nontrivial equations of motion or classical solutions. 
Nevertheless, the associated path integral
\begin{equation}
Z=\int {\cal D}[g] \, \mathrm{e}^{iS[g]},\;\;\;\;\; S=-\Lambda \int_M d^2x\, \sqrt{|g|} 
\label{contpi}
\end{equation}
gives rise to a nontrivial model of quantum geometry, which is an excellent testing ground for new gauge-invariant observables. 
Its nonperturbative evaluation beautifully illustrates the power of random-geometric methods, 
where identical building blocks are assembled into piecewise flat manifolds in all possible ways, subject to certain gluing rules.
They represent a regularized version of the configuration space of all geometries. 
In 2D CDT, the elementary building block is a flat Minkowskian isosceles triangle,
with one spacelike and two timelike edges of squared edge length $\ell_s^2\! =\! a^2$ and $\ell_t^2\! =\! -\alpha a^2$ respectively, where $a$ is 
the lattice spacing or UV cut-off and $\alpha\! >\! 0$ a positive constant (Fig.\ \ref{fig:tristrip}, left). 
These triangles (``two-simplices") are glued together pairwise along edges of matching length, such that the resulting triangulated geometry $T$ is a simplicial
manifold with a 
well-defined causal structure, amounting to a discrete version of global hyperbolicity. For compact spatial $S^1$-topology this is achieved by constructing each
$(1+1)$-dimensional geometry as a sequence of annular strips, labelled by a discrete ``proper time" parameter $t$ (Fig.\ \ref{fig:tristrip}, right). 

In terms of these ingredients, the formal path integral (\ref{contpi}) takes the explicit form 
\begin{equation}
Z_\mathrm{CDT}(\lambda) =\sum_T \frac{1}{C(T)}\, \mathrm{e}^{-S_\mathrm{CDT}(T)}, \;\;\;\;\; S_\mathrm{CDT} (T)=\lambda N_2(T),
\label{discpi}
\end{equation}
where we have used the Wick rotation of CDT (analytic continuation $\alpha\!\mapsto\! -\alpha$ in the lower-half complex plane \cite{Ambjorn2001,review1}) to obtain
a real partition function $Z_\mathrm{CDT}$, which is easier to work with computationally and will be the setting for the work presented here. 
In the Euclideanized path integral (\ref{discpi}), $\lambda$ is the bare, 
dimensionless cosmological constant,
multiplying the discrete volume of the triangulation $T$, given by the number $N_2(T)$ of triangles contained in it\footnote{In general, $N_k(T)$ denotes the number of
$k$-simplices in $T$.}, and $C(T)$ denotes the order of the automorphism group of $T$. 
The sum is over distinct Wick-rotated CDT geometries\footnote{Note that these triangulations are piecewise flat manifolds of Riemannian signature after
the analytic continuation, but still carry a memory of the original causal gluing rules.} and can be performed in the continuum limit $a\rightarrow 0$ while tuning $\lambda$ to its known
critical value $\lambda_c\! =\! \ln 2$ from above according to
\begin{equation}
\lambda \rightarrow \lambda_c +a^2 \Lambda+ O(a^3),
\label{lambda}
\end{equation}
where $\Lambda$ is the dimensionful renormalized cosmological constant \cite{Ambjorn1998}. Expectation values of geometric observables ${\cal O}(T)$ are
computed according to
\begin{equation}
\langle {\cal O}\rangle =\frac{1}{Z_\mathrm{CDT}}\sum_T \frac{1}{C(T)}\, {\cal O}(T)\, \mathrm{e}^{-S_\mathrm{CDT}(T)}
\label{expval}
\end{equation}
in the same limit. Sufficiently simple observables, like the spectral and Hausdorff dimension \cite{Ambjorn1998,spec2d} and the spectrum of the underlying 
quantum-mechanical Hamiltonian \cite{DiFrancesco,review1} can be computed analytically, but this has not yet been possible for the quantum Ricci curvature.
Following earlier work in 2D CDT on the average quantum Ricci curvature as a function of a local coarse-graining scale \cite{Brunekreef2021}, 
the so-called curvature profile \cite{Brunekreef2020}, we will investigate here its two-point correlator by a numerical analysis. 

\begin{figure}[t]
\centering
\includegraphics[width=0.5\textwidth]{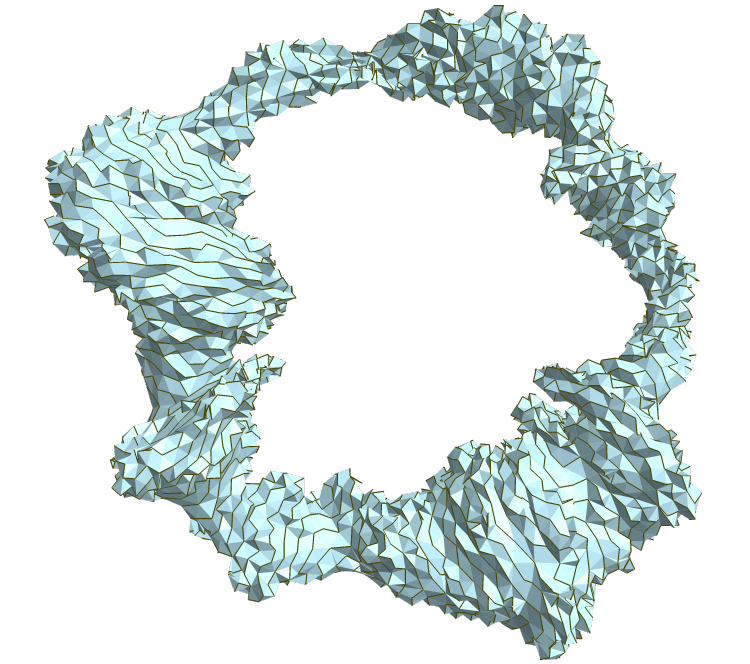}
\caption{Typical 2D toroidal CDT configuration at $N_2\! =\! 10$k, with spatial $S^1$-slices of fluctuating length (marked by dark lines) and time 
extension $t_\mathrm{tot}\! =\! 123$. Time is cyclically identified. 
}
\label{fig:cooltorus}
\end{figure}

For computational convenience, we will adopt several standard choices. We will cyclically identify the time direction, such that the topology of the geometries is that
of a two-torus $T^2\!\equiv\! S^1\!\times\! S^1$ (Fig.\ \ref{fig:cooltorus}). Each one-dimensional spatial universe at constant integer time $t\!\in\! [0,1,2,\dots,t_\mathrm{tot}]$,
where times 0 and $t_\mathrm{tot}$ are identified, consists of a closed chain of $\ell(t)\!\geq\! 3$ spacelike edges. 
The simulations will be performed at various fixed two-volumes $N_2\!\in\! [50\mathrm{k},300\mathrm{k}]$
and fixed numbers of time steps in such a way that 
the quotient $t_\mathrm{tot}^2/N_2\approx 0.16$ is approximately constant\footnote{More precisely, the combinations used are
$(N_2,t_\mathrm{tot})=(50\mathrm{k},89)$, (100k,126), (150k,155), (200k,179), (250k,200) and (300k,219).\label{fn4}},
leading to an average spatial volume $\bar{\ell}\! =\! N_2/(2 t_\mathrm{tot})$ in the range $\bar{\ell}\!\in\! [280,685]$.

Expectation values of geometric observables at fixed volume $N_2$ are given by
\begin{equation}
\langle {\cal O}\rangle_{N_2} =\frac{1}{Z_{N_2}}   \sum_{T|_{N_2}} \frac{1}{C(T)}\, {\cal O}(T),\;\;\;\;\;\; Z_{N_2}=  \sum_{T|_{N_2}} \frac{1}{C(T)},
\label{expfix}
\end{equation}
where the sums are over CDT configurations of volume $N_2$, and
the fixed-volume path integral $Z_{N_2}$ is related to the CDT path integral (\ref{discpi}) for cosmological constant $\lambda$ 
by a Laplace transform,
\begin{equation}
Z_\mathrm{CDT}(\lambda) =\sum_{N_2} \mathrm{e}^{-\lambda N_2} Z_{N_2}.
\label{}
\end{equation}
The behaviour of observables in the limit of infinite volume is extrapolated from sequences of measurements for 
finite, increasing $N_2$ by using finite-size scaling, a standard tool for analyzing statistical systems (see e.g.\ \cite{NB}).
We use a Monte Carlo Markov chain (MCMC) algorithm to generate sequences of independent CDT configurations, which allows
us to approximate the expectation values (\ref{expval}) of observables by importance sampling from the CDT ensemble.

One hallmark of gravitational path integrals formulated in terms of (causal) dynamical triangulations is the fact that no coordinates are needed to describe the
triangulated manifolds. The geometry of the individual building blocks is uniquely fixed by their geodesic edge lengths, and is complemented by connectivity data which specify 
the neighbourhood relations between pairs of building blocks. An important consequence in the present context is the absence of any nontrivial 
diffeomorphism- or coordinate-symmetry that needs to be taken into account when evaluating the path integral. This leads to an important simplification compared
to the situation in continuum path integrals, where such a gauge redundancy cannot be avoided. Note that during numerical simulations it becomes necessary 
to attach labels to the (sub-)simplices of a triangulation, which might be seen as analogous to using coordinates. However, these unphysical labels are 
discrete and the ensuing redundancy can be taken care of in a straightforward way in the computer algorithm
by working with labelled triangulations and appropriately taking relabelling multiplicities into account \cite{Ambjorn1997,review1}.

The computer code used in this work was written from scratch in the Rust programming language \cite{rust}. 
Source code and accompanying documentation are available via a GitLab repository \cite{gitlab}. 
It was tested thoroughly by measuring several standard quantities (vertex degree distribution of a CDT lattice, Hausdorff and spectral dimensions), 
and comparing them to the analytically known values\footnote{In the infinite-volume limit, the probability distribution of the vertex degree (coordination number) 
$c$ is given by $p[c]\! =\! \tfrac{c-3}{2^{c-2}}$ for $c\!\geq\! 4$ \cite{2dmatter}. The Hausdorff dimension $d_H$ and spectral dimension $d_S$ 
are both equal to 2 \cite{Ambjorn1998,spec2d}.}
as well as previous measurements, yielding compatible results.\footnote{We found a discrepancy between the measured and analytically known Hausdorff dimension, 
which was also the case in previous investigations, see \cite{vdduin} for further details.}

\begin{figure}[t]
\centering
\includegraphics[width=0.8\textwidth]{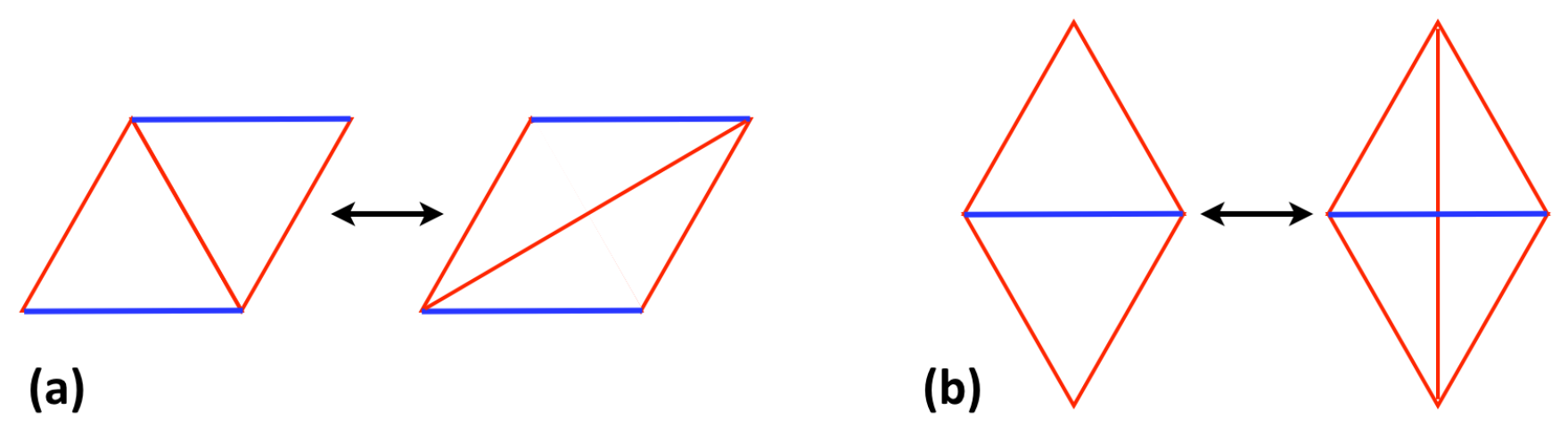}
\caption{The two standard Monte Carlo moves in 2D CDT, the flip move (a) and the 2-4 or 4-2 move (b), depending on the direction of the arrow.}
\label{fig:mcmoves}
\end{figure}

We will not describe all details of the Monte Carlo simulation, which do not deviate significantly from earlier implementations, and point 
the interested reader to reference \cite{vdduin}. 
The only novelty we introduced in the 2D simulations is a set of ergodic, \textit{volume-preserving} Monte Carlo moves, 
thereby avoiding the need to perform simulations in a small
window $[N_2\!-\!\Delta N_2,N_2\!+\!\Delta N_2]$ around a target volume $N_2$. Recall that the standard moves
to create a Markov chain $\{ T_0, T_1, T_2, \dots \}$ of CDT configurations are two types of local alterations of a triangulation, the flip move (Fig.\ \ref{fig:mcmoves}a)
and the 2-4 move together with its inverse 4-2 move (Fig.\ \ref{fig:mcmoves}b) \cite{Ambjorn2001}. The latter is not volume-preserving, since it adds or removes two triangles. 
We have substituted it by a nonlocal move that combines the application of a 2-4 move with the simultaneous application of a 4-2 move in a non-overlapping location on
the same triangulation $T_i$, and leaves $N_2$ invariant.\footnote{We thank T.\ Budd for suggesting this construction.} It is easy to show
that this new move together with the flip move is ergodic in the CDT ensemble of a given volume $N_2$ and number $t_\mathrm{tot}$ of spatial slices. 
A systematic discussion for 2D and 3D CDT of the Metropolis-Hastings algorithm, Monte Carlo moves, detailed balance conditions, optimal data storage and retrieval, 
and issues of tuning, thermalization and measurement can be found in \cite{simul}.

\section{Quantum Ricci curvature}
\label{sec:qrc}

Given the central importance of curvature in general relativity, one may wonder whether there is a notion of \textit{quantum curvature}
that can play an analogous role in nonperturbative quantum gravity. In view of the second-order nature of the classical Riemann tensor, it is
unclear a priori how to associate a local curvature with a nonsmooth space, like those occurring in the CDT path integral. 
To be useful from a physical point of view, such a notion should be well-defined and finite in 
a strongly quantum-fluctuating microscopic regime and at the same time relate to standard notions of curvature on macroscopic scales, where
geometry is smooth and classical. The quantum Ricci curvature (QRC) introduced in \cite{qrc1} has been shown to satisfy both criteria. 
It is a quasi-local quantity which relies on distance and volume measurements, but
does not require smoothness or the availability of tensor calculus. A comprehensive review of the QRC and its applications, and of other attempts to define curvature
in nonperturbative (C)DT quantum gravity can be found in \cite{curvsumm}; below we will give a compact summary of the properties of the QRC that are relevant to the
construction of the curvature correlator.

The QRC is based on the geometric insight that in a Riemannian space, in the presence of positive (negative) Ricci curvature, 
the distance -- appropriately defined -- between two sufficiently close and sufficiently small geodesic spheres is smaller (bigger) than the distance between their 
centres \cite{Ollivier2011}. This can be turned around to \textit{define} a generalized notion of Ricci curvature on
general, nonsmooth metric spaces \cite{ollivier}. Exploiting this idea, the QRC uses as a distance the 
\textit{average sphere distance} $\bar d$. For two $(D\! -\! 1)$-spheres $S^\delta_p$ and $S^\delta_{p'}$
of radius $\delta$ on a $D$-dimensional Riemannian manifold $(M,g_{\mu\nu})$, with centres $p$ and $p'$, it is defined as \cite{qrc1}
\begin{equation}
\bar{d}(S_p^{\delta},S_{p'}^{\delta})\! :=\!\frac{1}{\mathrm{vol}\,(S_p^{\delta})}\frac{1}{\mathrm{vol}\,(S_{p'}^{\delta})}
\int_{S_p^{\delta}}\! d^{D-1}\! q\, \sqrt{h} \int_{S_{p'}^{\delta}}\! d^{D-1}\! q'\, \sqrt{h'}\ d_g(q,q'),\;\;\; d_g(p,p')\! =\! \varepsilon,
\label{sdist}
\end{equation}
which is simply the distance average of all point pairs $(q,q')\!\in\! S^\delta_p \times S^\delta_{p'}$.
In relation (\ref{sdist}), $h$ and $h'$ denote the determinants of the induced metrics on the two spheres,
$\mathrm{vol}(S^\delta)$ denotes the volume of the sphere with respect to the corresponding induced metric, 
and $d_g(q,q')$ is the geodesic distance of the points $q$ and $q'$ with respect to the metric $g$.

Our application will involve a simplicial implementation of (\ref{sdist}) for $D\! =\! 2$ and for the case $\varepsilon\! =\! 0$, such that $p'\! =\! p$ and
the two spheres coincide. The latter is a convenient choice since we will only be interested in correlators of
the direction-independent (quantum) Ricci scalar, and not the full quantum Ricci curvature, although for simplicity we will still refer to it as ``QRC" in what follows.
For vanishing $\varepsilon$, (\ref{sdist}) measures the distance of the sphere $S^\delta_p$ to itself. To exhibit the relation of this quantity 
with the local Ricci scalar $R(p)$ in the continuum case, 
we divide the expression for $\bar d$ by $\delta$ to obtain the \textit{normalized average sphere distance}. Evaluating it at a point $p$ of a 2D Riemannian
manifold as a power series in $\delta$ yields
\begin{equation}
\bar{d}(S_p^\delta,S_p^\delta)/\delta =\frac{4}{\pi}-\frac{\delta^2}{9\pi} R(p)+O(\delta^3)\approx 1.273-0.0354\,\delta^2 R(p)+O(\delta^3),
\label{nasd}
\end{equation}
as may be verified, say, by a computation in Riemann normal coordinates \cite{brewin}.
This expression is valid for $\delta$ sufficiently small such that the geodesics involved in the measurement of pairwise distances exist and are unique.
The particular value of the constant term is related to our choice of sphere distance \cite{qrc1}. Apart from this finite rescaling of the constant term, 
the expansion (\ref{nasd}) captures our earlier statement about the geometric insight underlying the QRC,
since the coefficient of the lowest-order ``correction" term proportional to $\delta^2$ to the constant term is a negative number times the Ricci scalar. 
Higher-order terms in the $\delta$-expansion (\ref{nasd}) depend on the Ricci scalar and its covariant derivatives. In other words, for vanishing Ricci scalar
the function $\bar{d}/\delta$ will be constant in $\delta$, while for positive (negative) Ricci scalar it will initially decrease (increase) as a function of $\delta$. 

In terms of the normalized averaged sphere distance (\ref{nasd}), the \textit{quantum Ricci curvature} $K_q(p;\delta)$ is defined by
\begin{equation}
\bar{d}(S_p^\delta,S_p^\delta)/\delta =: c_q (1-K_q(p;\delta)),
\label{qrc}
\end{equation}
where $c_q\! :=\! \lim_{\delta\rightarrow 0} \bar{d}/\delta$ in the smooth case and $K_q$ captures the $\delta$-dependent part of the left-hand side. 
The subscript ``$q$" stands for ``quantum", although the context at this stage is not (yet) the quantum theory. 
On smooth manifolds, $c_q$ is a constant satisfying $0\! <\! c_q\! <\! 2$ which only depends on the dimension. 
However, the main motivation for the definition (\ref{qrc}) is that it allows us to associate a curvature with more general metric spaces, and
piecewise flat triangulated spaces in particular. On such a triangulation, distances and volumes come in discrete lattice units. On the 2D CDT configurations
considered here, we will use the \textit{geodesic link distance} $d(q,p)$ between pairs $(q,p)$ of lattice vertices, which is given by the number of edges in the shortest path 
along edges between $q$ and $p$.\footnote{Note that without loss of generality the triangular building blocks after the Wick rotation can be taken to be equilateral,
implying that all edges have length $a$.} 
The geodesic sphere $S_p^\delta$ of radius $\delta$ with central vertex $p$ is defined as the set of all vertices $q$ that have link distance $\delta$
to $p$, and the discrete volume $N_0(S_p^\delta)$ of $S_p^\delta$ is the number of vertices contained in it. 
In terms of these ingredients, the normalized average sphere distance for $\varepsilon\! =\! 0$ becomes\footnote{Since the local quantities we will
consider are mainly associated with vertices of the triangulation, any averages will typically involve the number $N_0(T)$ of vertices. On a torus, this is related
to the number $N_2(T)$ of triangles in $T$ by $N_2(T)\! =\! 2 N_0(T)$.}
\begin{equation}
\bar{d}(S_p^{\delta},S_{p}^{\delta})/\delta  =\frac{1}{N_0(S_p^\delta)^2} \sum_{q\in S_p^\delta}\sum_{q'\in S_p^\delta} d(q,q')/\delta .
\label{discnasd}
\end{equation}
As one would expect, this quantity suffers from lattice artefacts for small $\delta$, i.e.\ a behaviour that is sensitive to the way the discretization has
been set up. Inves\-ti\-ga\-tions of the QRC on various classical and dynamically triangulated lattices in 2D have consistently found significant lattice artefacts 
in a region $\delta\!\lesssim\! 5$ \cite{qrc1,qrc2,Brunekreef2021}. Since we are interested in the
continuum limit of these regularized models, we should discard data points from this region. 
We also need to define a suitable lattice analogue of the limit $\delta\!\rightarrow\! 0$ when extracting the QRC via relation (\ref{qrc}), as well as
take into account that the value of $c_q$ is not universal, but depends on the discretization \cite{qrc1}. 

For nonclassical triangulations, which are not designed to approximate a specific classical manifold, we do not know the functional form in $\delta$ of 
the normalized average sphere distance $\bar{d}(S_p^\delta,S_p^\delta)/\delta$ a priori. In particular, there is no guarantee that in a $\delta$-window
outside the region of lattice artefacts it can be fitted to a curve of the form $\gamma(\delta)\! =\!\alpha\! -\! \beta \delta^2\! +h.o.$, for constants $\alpha$, $\beta$, 
as a function of the distance $\delta$ and in generic points $p$.
If such a fit is possible, it constitutes nontrivial evidence for the presence of a positive, negative or vanishing curvature compatible with that of a continuum manifold,
at least when the distance $\delta$ scales canonically, i.e.\ behaves like a length.\footnote{In quantum applications, this will also involve a finite-size scaling analysis and
a substitution of dimensionless lattice quantities by their dimensionful counterparts.} 
If such a fit is not possible, a positive or negative deviation of $\bar{d}/\delta$ from constancy according to our definition (\ref{qrc}) still gives rise to a 
nonvanishing quantum Ricci curvature, characterizing the metric space in question, 
but its geometric (continuum) interpretation may be less clear-cut. This latter situation appears to be realized
in 2D CDT \cite{Brunekreef2021}, as we will discuss later in this section. This is not completely surprising, since 2D quantum gravity is a pure 
quantum theory; the Einstein-Hilbert action in 2D does not possess nontrivial geometric solutions, which means that there are no classical spacetimes that 
can be obtained from the quantum theory in some classical limit.  

For use in the quantum theory, we must construct diffeomorphism-invariant observables from these ingredients. The most straightforward way of doing this is
to take the average $\bar{d}_\mathrm{av}$ over the triangulation $T$ of the average sphere distance. Its normalized version gives rise to what has been dubbed 
the \textit{curvature profile} of $T$ \cite{Brunekreef2020}, 
\begin{equation}
\bar{d}_\mathrm{av}(\delta)/\delta:=\frac{1}{N_0(T)}\sum_{p\in T} \bar{d}(S_p^{\delta},S_{p}^{\delta})/\delta,
\label{cprof}
\end{equation}
which can be used to extract an average QRC $K_\mathrm{av}(\delta)$ via
\begin{equation}
\bar{d}_\mathrm{av}(\delta)/\delta =: c_\mathrm{av}(1-K_\mathrm{av}(\delta)),
\label{cprofav}
\end{equation}
where the constant $c_\mathrm{av}$ should be fixed appropriately, e.g.\ by the value of $\bar{d}_\mathrm{av}/\delta$ just outside the region of lattice artefacts.

The expectation value of the curvature profile (\ref{cprofav}) has been investigated for Euclidean DT quantum gravity in 2D \cite{qrc2}, and for Lorentzian CDT quantum gravity in
2D \cite{Brunekreef2021} and 4D \cite{qrc3}. In all cases the standard
QRC, based on the direction-dependent average sphere distance (\ref{sdist}) with $\varepsilon\! =\! \delta$ was used, followed by a directional averaging. 
We have repeated the measurements of \cite{Brunekreef2021} for 2D CDT, as a further cross-check of our code, and found excellent agreement, 
as long as $\delta$ is sufficiently small to not be affected by finite-size effects\footnote{and taking into account that a different method of sampling pairs of
spheres was used in \cite{Brunekreef2021}, see Appendix B for further discussion}. The latter arise because of the compactness of the triangulations in both the time and
spatial directions. When the spheres involved in the sphere distance measurements become sufficiently large, there will be
point pairs $(q,q')$ contributing to (\ref{discnasd}) whose associated geodesic -- which we have defined as the \textit{globally} shortest path between $q$ and $q'$ -- will run around
the ``back side" of the torus, i.e.\ take a ``topological shortcut" that would not be available if the torus was cut open to remove the compactification. 
For even larger $\delta$, the spheres themselves will eventually start wrapping around the compact directions, giving rise to self-overlaps. 

Since we are primarily interested in local curvature properties, we want to control and eliminate such global, finite-size effects as much as possible. 
This is not completely straightforward in the case at hand because the geometry fluctuates: although the average size $\bar{\ell}$ of spatial slices is determined
by the two-volume $N_2$ and time extension $t_\mathrm{tot}$, fluctuations $\Delta\ell(t)$ are large \cite{Ambjorn1998}, which means that the length of the shortest closed geodesic 
winding once around the spatial direction depends on the configuration $T$ and will typically be much smaller than $\bar{\ell}$. One could in principle keep track
of the nature of the geodesic of each point pair $(q,q')$, but this is computationally costly. Instead, the strategy developed in \cite{Brunekreef2021} was to
establish an upper bound $\delta_\mathrm{max}$ on the sphere radii $\delta$ for given $N_2$ and $t_\mathrm{tot}$, such that for $\delta\!\leq\!\delta_\mathrm{max}$
occurrences of topological shortcuts are sufficiently rare to not distort measurements of the average sphere distance significantly. 

\begin{figure}[t]
\centering
\includegraphics[width=0.7\textwidth]{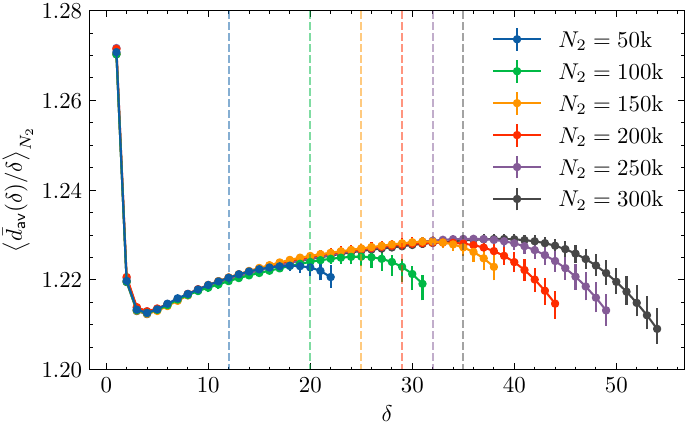}
\caption{
    Curvature profile $\langle \bar{d}_\mathrm{av}(\delta)/\delta \rangle_{N_2}$ for $\varepsilon \! =\! 0$ measured in 2D CDT ensembles of fixed volume
    $N_2\!\in\! [50\mathrm{k},300\mathrm{k}]$. The vertical dashed lines
    indicate thresholds $\delta_\mathrm{max}$ for the sphere radii, below which finite-size effects are negligible for the corresponding volume $N_2$. 
    (Here and in subsequent figures, discrete data points are connected by straight line segments for easier visualization.)
}
\label{fig:curvprofile}
\end{figure}

In the present work, we have used a slight variant of the criterion in \cite{Brunekreef2021} to find a set $\{\delta_\mathrm{max} \}$ of such thresholds,
see \cite{vdduin} for details.
For lattice volumes $N_2\!\in\! [50\mathrm{k},300\mathrm{k}]$, they lie in the range $\delta_\mathrm{max}\!\in\! [12, 35]$. 
In the plots of the expectation value $\langle \bar{d}_\mathrm{av}(\delta)/\delta \rangle_{N_2}$ of the curvature profile (or normalized average sphere distance) 
shown in Fig.\ \ref{fig:curvprofile}
they are indicated by vertical dashed lines. The measurements of the curvature correlators presented in Sec.\ \ref{sec:correl} 
below use only values for $\delta$ that lie well below these thresholds.

The curvature profile of Fig.\ \ref{fig:curvprofile} we have measured for $\varepsilon\! =\! 0$ 
is qualitatively very similar to the standard case of the QRC with $\varepsilon\! =\! \delta$ investigated previously \cite{Brunekreef2021},
and consistent with these earlier findings. 
The region $\delta\!\lesssim\! 5$ shows a rapid fall-off that is virtually identical for all volumes, a typical lattice artefact. Beyond it, each of the
curves increases gently until eventually dropping off again, the latter being a known finite-size effect due to the underlying compactified spacetime topology. Within error bars,
the curves coincide within their relevant $N_2$-dependent ranges $\delta\!\in\! [5,\delta_\mathrm{max}]$. The joint enveloping curve of the curvature profile increases, 
with a decreasing slope for growing $\delta$, similar to what was observed in \cite{Brunekreef2021}, and indicating the presence of
negative curvature. However, in line with our remarks earlier in this section the function $\langle \bar{d}_\mathrm{av}(\delta)/\delta \rangle_{N_2}$ 
beyond $\delta\!=\! 5$ appears to be concave. It therefore cannot be fitted to a quadratic
deviation from constancy of the form $-\beta\delta^2$, $\beta\! <\! 0$, which would indicate compatibility with a negative average (continuum) Ricci scalar, 
because this is associated with a convex curvature profile. 
We conclude that 2D CDT exhibits the presence of a nonvanishing, negative average QRC, as captured by relation (\ref{cprofav}).\footnote{Note that this does not contradict 
the Gauss-Bonnet theorem, since there is no 2D classical manifold or geometry associated with the continuum limit of 2D CDT. 
Alternative notions of (quantum) curvature in 2D, including Regge's deficit-angle curvature \cite{regge} or variants of the QRC 
\cite{altcurv}, may behave differently.}   
This motivates the study of curvature correlators based on the QRC, which we will turn to next.

\section{Connected correlators}
\label{sec:correl}

The definition and implementation of geometric, invariant two-point functions in quantum gravity comes with some subtleties and ambiguities, 
which we will discuss before presenting the outcomes of our measurements for the curvature correlators. They are related to the fact that geometry fluctuates and 
there is no a priori fixed background geometry. The resulting nonuniqueness of the correlation functions concerns their normalization, the role of
manifold versus ensemble average and the subtraction prescription to obtain connected correlators. Different choices lead to inequivalent
prescriptions of correlators at the level of the regularized theory. 
Whether or not they also lead to physically inequivalent results in the continuum limit 
has to be investigated in detail for a given observable and a given quantum gravity model. As we will see below, we found good evidence in the 
case at hand that 
the differences between these choices become negligible for large volumes $N_2$ and distances $r$, but this was by no means a foregone conclusion.

\subsection{Classical preliminaries}
\label{subsec:class}

We will temporarily revert to a continuum formulation in terms of a two-dimen\-sio\-nal compact Riemannian manifold $(M,g_{\mu\nu})$ to introduce
some key elements of the correlator construction.
Given two local scalar functions ${\cal O}_1(x)$, ${\cal O}_2(x)$ of the metric $g_{\mu\nu}$, we can define the diffeomorphism-invariant quantity
\begin{equation}
G_g[{\cal O}_1,{\cal O}_2](r)\!\! :=\!\!\! \int_M\! d^2x \sqrt{g(x)}\int_M\! d^2y \sqrt{g(y)}\, {\cal O}_1(x){\cal O}_2(y)\, \delta (d_g(x,y)-r),
\label{gclass}
\end{equation}
which may be thought of as a ``classical two-point function" of ${\cal O}_1$ and ${\cal O}_2$, depending on the geodesic distance $r$ and the geometry $g$
(metric modulo diffeomorphisms). The integrations over the locations of the two points $x$ and $y$ are necessary to obtain a
diffeomorphism-invariant quantity, which is also required to make any subsequent ensemble average meaningful.
We are not suggesting that (\ref{gclass}) is an interesting observable from the point of view of classical general relativity, but only that it is a useful 
ingredient of the quantum theory, where we will consider expectation values of invariant observables in the (nonperturbative) vacuum state of the theory.
By construction, $G_g[{\cal O}_1,{\cal O}_2](r)$ is symmetric in its two functional arguments. 
Since the manifold is assumed to have a finite volume 
\begin{equation}
V_g=\int_M d^2x\sqrt{g},
\label{volm}
\end{equation}
(\ref{gclass}) must vanish for sufficiently large $r$, whenever there are no more point pairs in $M$ that are sufficiently far apart. 
Noting that the constant function ${\cal O}(x)\! =\! 1$ is a special case, one can construct two classical correlators that will
play an important role in what follows. For two insertions of the unit function one obtains
\begin{eqnarray}
G_g[1,1](r)\!\!\! &=&\!\!\!  \int_M d^2x \sqrt{g(x)}\int_M d^2y \sqrt{g(y)}\, \delta (d_g(x,y)-r)\nonumber \\
\!\!\! &=&\!\!\! \int_M d^2x \sqrt{g(x)}\, \mathrm{vol}(S_x^r) = V_g\, \overline{\mathrm{vol}(S^r)}|_g,
\label{ones}
\end{eqnarray}
which is a measure of the number of all point pairs with mutual distance $r$ in the geometry $g$.
The overbar in (\ref{ones}) denotes the diffeomorphism-invariant manifold average,
\begin{equation}
\overline{\cal O}|_g:=\frac{1}{V_g}\int_M d^2x\sqrt{g(x)}\, {\cal O}(x), 
\label{manav}
\end{equation}
where our notation highlights its $g$-dependence for added clarity.
If one of the functions is the unit function, we have
\begin{eqnarray}
G_g[{\cal O},1](r)\equiv G_g[1,{\cal O}](r)\!\!\! & =&\!\!\! \int_M d^2x \sqrt{g(x)}\, \int_M d^2y \sqrt{g(y)}\, {\cal O}(x)\; \delta (d_g(x,y)-r)\nonumber\\
\!\!\! &=&\!\!\! \int_M d^2x \sqrt{g(x)}\, {\cal O}(x)\, \mathrm{vol}(S_x^r),
\label{oneando}
\end{eqnarray}
which, up to an overall factor of $V_g$, is the manifold average of the product of ${\cal O}(x)$ with a non\-local weight factor equal to the size of the sphere $S_x^r$
of radius $r$ centred at $x$. The source of this nonlocality is the geodesic distance $d_g(x,y)$ between point pairs $(x,y)$, which depends nonlocally on the metric. 

A ``connected" version of the classical two-point function (\ref{gclass}) is obtained by substituting the functional arguments ${\cal O}_i(x)$ by the deviations from 
their means $\overline{\cal O}_i|_g$, yielding
\begin{eqnarray}
&&\hspace{-1.4cm} G^c_g[{\cal O}_1,{\cal O}_2](r):=\nonumber \\
&&\hspace{-0.7cm} \int_M\! d^2x \sqrt{g(x)}\int_M\! d^2y \sqrt{g(y)}\, ({\cal O}_1(x)\! -\!\overline{\cal O}_1|_g)( {\cal O}_2(y)\! -\!\overline{\cal O}_2|_g)\, \delta (d_g(x,y)-r),
\label{gconn}
\end{eqnarray}
which again is of the form of a diffeomorphism-invariant twofold manifold integral. 
In terms of the previous two-point functions (\ref{gclass}), (\ref{ones}) and (\ref{oneando}), this expression can be written as 
\begin{eqnarray}
&&\hspace{-1.4cm}G^c_g[{\cal O}_1,{\cal O}_2](r) =\nonumber\\ 
&&\hspace{-0.8cm} G_g[{\cal O}_1,{\cal O}_2](r)\! -\!\overline{\cal O}_1|_g G_g[{\cal O}_2,1](r)\! -\! \overline{\cal O}_2|_g G_g[{\cal O}_1,1](r)\! +\!
\overline{\cal O}_1|_g \overline{\cal O}_2|_g G_g[1,1](r).
\label{gconnexp}
\end{eqnarray}
Note in particular the nature of the ``mixed" terms depending on $G_g[{\cal O}_i,1](r)$, which depend nontrivially on the metric, as we have
pointed out above. This dependence becomes trivial when the metric is constant, in the sense of being maximally symmetric, i.e.\ having a maximal number of Killing vectors. 
Specializing to the two-dimensional case at hand, maximal symmetry of the 2D geometry $g$ on $M$ implies that $\mathrm{vol}(S^r_x)$ does not depend
on $x$, and $G_g[{\cal O}_i,1]$ factorizes according to
\begin{equation}
G_g[{\cal O}_i,1](r)\! =\! \mathrm{vol}(S^r)|_g\!\! \int_M\! d^2x \sqrt{g(x)}\, {\cal O}_i(x)\! =\! \mathrm{vol}(S^r)|_g V_g\, \overline{\cal O}_i|_g \;\;\; [g\; \mathrm{max.}\; \mathrm{symmetric}],
\label{devel1}
\end{equation}
where $\mathrm{vol}(S^r)|_g\! =\! \overline{\mathrm{vol}(S^r)}|_g$.
As a consequence, the last three terms on the right-hand side of (\ref{gconnexp}) are all equal (up to their signs) and the expression simplifies to 
\begin{equation}
G^c_g[{\cal O}_1,{\cal O}_2](r)\!=\! G_g[{\cal O}_1,{\cal O}_2](r)\! -\! V_g\, \mathrm{vol}(S^r)|_g  \overline{\cal O}_1|_g\, \overline{\cal O}_2|_g 
\;\;\;\; [g\; \mathrm{max.}\; \mathrm{symmetric}].
\label{devel2}
\end{equation}
A further simplification occurs when one considers normalized versions of these classical expressions, which we will denote by $\tilde{G}_g$ and $\tilde{G^c_g}$. 
A natural normalization for given $r$ is obtained by dividing by expression (\ref{ones}),
\begin{equation}
\tilde{G}_g[{\cal O}_1,{\cal O}_2](r) :=  \frac{G_g[{\cal O}_1,{\cal O}_2](r)}{G_g[1,1](r) },
\label{defnorm}
\end{equation}
and similarly for $\tilde{G}^c_g$.
For maximally symmetric $g_{\mu\nu}$ this yields the simple form                                                                        
\begin{equation}
\tilde{G}^c_g[{\cal O}_1,{\cal O}_2](r)\! =\!   \tilde{G}_g[{\cal O}_1,{\cal O}_2](r) -
\overline{\cal O}_1|_g \, \overline{\cal O}_2|_g \;\;\;\;\; \;\;\;\;[g\; \mathrm{max.}\; \mathrm{symmetric}].                            
\label{connsymm}
\end{equation}
                                                                                                                                                                                                             
\subsection{Correlators in 2D CDT quantum gravity}
\label{subsec:cdt}

After these preliminary classical considerations we now turn to the definition of genuine two-point correlation functions in the (fixed-volume) quantum ensemble of 2D CDT.
We will largely adhere to the notation of the previous subsection, where $r$ will now refer to the geodesic link distance, the delta function of $r$ is replaced by a Kronecker delta, and the dependence on a geo\-metry $g$ is replaced
by the dependence on a triangulated CDT configuration $T$. (Quasi-)local quantities $\cal O$ will be associated with vertices $x\!\in\! T$, 
where we associate with each vertex a volume weight 1. As a consequence, (simplicial) manifold averages are given by
\begin{equation}
\overline{\cal O}|_T=\frac{1}{N_0(T)}\sum_{x\in T} {\cal O}(x),
\label{tav}
\end{equation}
and the lattice implementation of the classical two-point function (\ref{gclass}) has the form
\begin{equation}
G_T[{\cal O}_1,{\cal O}_2](r)\! :=\! \sum_{x,y\in T} {\cal O}_1(x){\cal O}_2(y)\, \delta_{d(x,y),r}.
\label{tclass}
\end{equation}
We will always compute expectation values of observables in the fixed-volume ensemble, as defined by (\ref{expfix}).
For simplicity of notation we will drop the subscript $N_2$, i.e.\ henceforth $\langle \cdot \rangle_{N_2} \!\equiv\! \langle \cdot\rangle$.
With this understanding, a natural definition of an (unnormalized) connected two-point correlator, inspired by (\ref{gconn}) and (\ref{gconnexp}), is
\begin{eqnarray}
\langle G^c_T[{\cal O}_1,{\cal O}_2](r)\rangle \!\! \!&=& \!\!\!
\langle G_T[{\cal O}_1,{\cal O}_2](r)\rangle  -\langle \overline{\cal O}_1|_T\, G_T[{\cal O}_2,1](r)\rangle\! \nonumber\\
&&\hspace{-0.3cm} - \langle\overline{\cal O}_2|_T\, G_T[{\cal O}_1,1](r)\rangle +
\langle \overline{\cal O}_1|_T\, \overline{\cal O}_2|_T\, G_T[1,1](r)\rangle.
\label{qugconnexp}
\end{eqnarray}
Due to its quantum nature, the physical interpretation of this correlator is quite different from that of
its classical precursor (\ref{gconn}). If one has only worked with correlation functions of nongravitational quantum fields on a fixed Minkowskian or Euclidean background,
an unexpected feature of (\ref{qugconnexp}) may be the presence of nontrivial correlations involving the unit operator, a quantity that does not depend on the dynamical 
field of the theory! The generic occurrence of such correlations was already noted in earlier treatments of Euclidean lattice gravity \cite{bakkersmit,conncorr}. 
However, a closer examination reveals that their origin is not the insertion of the unit operator at $x$ or $y$, but the 
$r$-dependence of $G_T[\cdot,\cdot](r)$. The distance $r$ refers to the length of the shortest geodesic between $x$ and $y$ and therefore depends on the metric in a nonlocal way.
In other words, the so-called ``two-point function" $\langle G_T[{\cal O}_1,{\cal O}_2](r)\rangle$ has a \textit{three}fold dependence on geometry, through the local geometric operators 
${\cal O}_1$, ${\cal O}_2$ \textit{and} through the nonlocal distance operator $r$ that appears as an argument of the Kronecker delta.
Quantum fluctuations of the geometry affect all of them, generally giving rise to 
nontrivial correlations. This is an intrinsic feature of correlators in quantum gravity, whether treated perturbatively or nonperturbatively, including correlators of quantum
matter coupled to quantum gravity. In all cases,
diffeomorphism invariance requires that correlators should not depend on a coordinate, but on a physical distance, which in turn implies a nontrivial and nonlocal
dependence on the underlying quantized metric field. This is most transparent in an approach like ours, where the correlation functions depend explicitly on the 
geodesic distance, as opposed to, say, gauge-fixed formulations.\footnote{For examples of continuum treatments of gravitational correlators depending on geodesic distances, 
see \cite{Modanese1994,Froeb2017} and references therein.} 

Furthermore, the last three terms on the right-hand side of (\ref{qugconnexp}) also involve correlations between configuration averages $\overline{\cal O}_i|_T$ 
and either $G_T[{\cal O}_i,1](r)$ or $G_T[1,1](r)$. Since all of these quantities are already $T$-averaged individually, one would expect their 
correlations to be much smaller than those captured by the expectation values $\langle G_T[\cdot,\cdot](r)\rangle$. 
It turns out that eliminating these correlations, by substituting $\langle \overline{\cal O}_1|_T\, G_T[{\cal O}_2,1](r)\rangle\! \longrightarrow\!
\langle \overline{\cal O}_1|_T\rangle \langle G_T[{\cal O}_2,1](r)\rangle$ etc., is equivalent to using
a variant of the connected correlator (\ref{qugconnexp}), which is equally well motivated a priori. 
It is obtained by substituting the $T$-averaged mean values $\overline{\cal O}_i|_T$, whose continuum version we subtracted in the classical formula (\ref{gconn}),
by their quantum (=ensemble) means $\langle \overline{\cal O}_i|_T\rangle$, resulting in
\begin{eqnarray}
\hspace{-0.3cm}\langle G^c_{T,q}[{\cal O}_1,{\cal O}_2](r)\rangle \!\! \!&=& \!\!\!
\langle G_T[{\cal O}_1,{\cal O}_2](r)\rangle  -\langle \overline{\cal O}_1|_T\rangle\langle G_T[{\cal O}_2,1](r)\rangle\! \nonumber\\
&&\hspace{-0.3cm} - \langle\overline{\cal O}_2|_T\rangle\langle G_T[{\cal O}_1,1](r)\rangle +
\langle \overline{\cal O}_1|_T\rangle \langle \overline{\cal O}_2|_T\rangle\langle G_T[1,1](r)\rangle,
\label{qugconnexp1}
\end{eqnarray}
where the subscript $q$ stands for ``quantum".
There is yet another variant that has been proposed in the literature \cite{bakkersmit,bialas}, namely to use as a mean the ensemble-averaged
$r$-dependent normalized, weighted averages
\begin{equation}
\langle \tilde{G}_T[{\cal O}_i,1](r)\rangle :=\bigg\langle \frac{G_T[{\cal O}_i,1](r)}{G_T[1,1](r)}\bigg\rangle =\bigg\langle \frac{\sum_{x,y\in T} {\cal O}_i(x)\, \delta_{d(x,y),r} }{ \sum_{x,y\in T}  \delta_{d(x,y),r}}\bigg\rangle,
\label{gtav}
\end{equation}
which amounts to substituting all instances of $\langle \overline{\cal O}_i|_T\rangle$ in (\ref{qugconnexp1}) by $\langle \tilde{G}_T[{\cal O}_i,1](r)\rangle$. 
We will not consider this prescription further here, but we have verified that the
normalized version of the resulting connected curvature correlator away from the regions with discretization and finite-size effects
leads to results very similar to those for the normalized version of the correlator (\ref{qugconnexp1}) \cite{vdduin}.

Turning next to the normalization of the connected correlators, a natural normalization for a given geometry and distance $r$ we have already used in
(\ref{defnorm}) and (\ref{gtav}) is a division by $G_g[1,1](r)$ and $G_T[1,1](r)$ respectively.
It has two different quantum implementations, depending on whether we normalize before or after taking the
ensemble average. If we normalize the correlator (\ref{qugconnexp}), this leads to the two options
\begin{equation}
{\cal G}[{\cal O}_1,{\cal O}_2,r] :=\langle \tilde{G}^c_T[{\cal O}_1,{\cal O}_2](r)\rangle =\bigg\langle \frac{G^c_T[{\cal O}_1,{\cal O}_2](r)}{G_T[1,1](r)}\bigg\rangle
\label{opt1}
\end{equation}
and
\begin{equation}
{\cal G}[{\cal O}_1,{\cal O}_2,r]' := \frac{\langle G^c_T[{\cal O}_1,{\cal O}_2](r)\rangle }{ \langle G_T[1,1](r)\rangle}
\label{opt2}
\end{equation}
respectively. If we take the correlator (\ref{qugconnexp1}) as our starting point, we obtain
\begin{equation}
{\cal G}_q[{\cal O}_1,{\cal O}_2,r] :=\langle \tilde{G}^c_{T,q}[{\cal O}_1,{\cal O}_2](r)\rangle =\bigg\langle \frac{G^c_{T,q}[{\cal O}_1,{\cal O}_2](r)}{G_T[1,1](r)}\bigg\rangle
\label{qopt1}
\end{equation}
and
\begin{equation}
{\cal G}_q[{\cal O}_1,{\cal O}_2,r]' := \frac{\langle G^c_{T,q}[{\cal O}_1,{\cal O}_2](r)\rangle }{ \langle G_T[1,1](r)\rangle}
\label{qopt2}
\end{equation}
as possible choices. Our general expectation is that 
they will not differ much in the parameter region where the measurements are not subject to significant short-distance discretization artefacts or finite-size effects.
We have already commented briefly on the two different mean subtractions underlying the difference between (\ref{opt1}) and (\ref{qopt1}) and between
(\ref{opt2}) and (\ref{qopt2}), corresponding to subtracting the mani\-fold mean $\overline{\cal O}_i|_T$ and its ensemble average $\langle\overline{\cal O}_i|_T\rangle$
respectively when defining the connected correlator. However, for any (quasi-)local quantity ${\cal O}$, we expect the amount of self-averaging (on the same geometry)
to increase as the geometries $T$
grow in volume, such that the manifold averages approach the ensemble average ever more closely. Of course, such assumptions
need to be checked in explicit situations, especially when long-range correlations may be present. Even if certain quantities can be shown to become
independent in an infinite-volume limit, we are working with finite volumes where different prescriptions may differ appreciably. From a practical point of view, some 
choices of correlators may be less sensitive to finite-size effects than others, and therefore preferable. We have run explicit tests of the two different
subtraction prescriptions for the normalized correlators of the local vertex degree or coordination number $c(x)$ (the number of triangles meeting at the vertex $x$) 
and the average sphere distance $\bar{d}$ and did not observe any differences within measuring accuracy.  

\begin{figure}[t]
\centering
\includegraphics[width=0.8\textwidth]{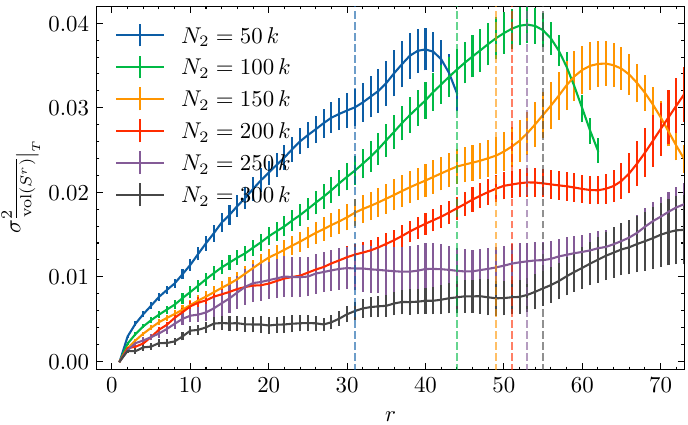}
\caption{Relative deviation of the manifold average of the sphere volume $\mathrm{vol}(S^r)$ from its ensemble average, as a function
of the sphere radius $r$, for volumes $N_2\!\in [50\mathrm{k},300\mathrm{k}]$. 
Vertical dashed lines indicate the upper end $r_\mathrm{max}$ of the trusted $r$-range (cf.\ Appendix A). The relatively large error bars come 
from using only about 100 configurations.}
\label{fig:deviationsv}
\end{figure}

The situation regarding the difference between the two normalizations, i.e.\ between (\ref{opt1}) and 
(\ref{opt2}) and between (\ref{qopt1}) and (\ref{qopt2}), is similar. Recalling that $G_T[1,1](r)\! =\! N_0(T)\overline{\mathrm{vol}(S^r)}|_T$, we see that
this difference depends on the variance
\begin{equation}
\sigma^2_{\overline{\mathrm{vol}(S^r)}|_T}\! =\! \langle \overline{\mathrm{vol}(S^r)}|_T{}^2 \rangle - \langle \overline{\mathrm{vol}(S^r)}|_T \rangle^2 .
\label{varvol}
\end{equation}
For $\sigma^2_{\overline{\mathrm{vol}(S^r)}|_T}\! =\! 0$, the two normalizations would be identical. We have verified that the variance (\ref{varvol})
decreases as a function of volume, for all distances $r$. Fig.\ \ref{fig:deviationsv} shows the square root of the variance relative to the ensemble mean 
$\langle \overline{\mathrm{vol}(S^r)}|_T \rangle$ of the sphere volume. 
Within error bars, this quantity decreases monotonically for all radii $r$, from the smallest volume $N_2\! =\! 50$k (top curve) to the largest one
$N_2\! =\! 300$k (bottom curve). For the largest volume, the relative deviation lies below or around half a percent in the trusted $r$-range, which is indeed small.
In addition, we have verified explicitly that the measured normalized connected correlators for the coordination number and the sphere distance for both
normalization prescriptions overlap within statistical error bars. 

To summarize, for the measurement of curvature correlators in 2D CDT, all variants (\ref{opt1})--(\ref{qopt2})
presented above seem to give equivalent results.
In our quantitative analysis below, we have made the specific choices to normalize the correlators by $G_T[1,1](r)$ \textit{before} taking the ensemble average, and
to always use a subtraction by the ensemble average to obtain connected correlators, which corresponds to option (\ref{qopt1}).

\begin{figure}[t]
\centering
\includegraphics[width=0.6\textwidth]{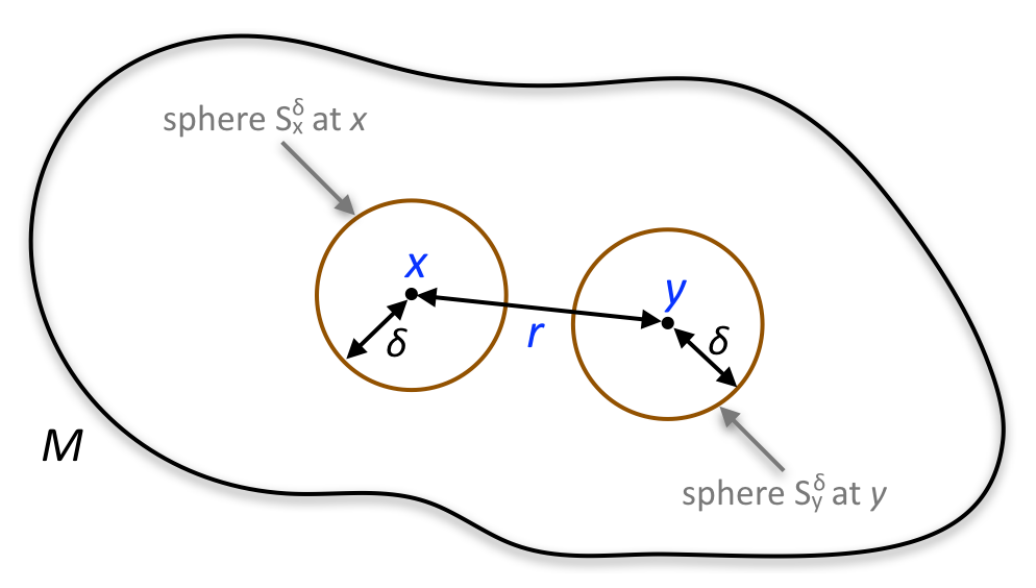}
\caption{Schematic illustration of the contribution of an individual point pair $(x,y)$ at geodesic distance $r$ to the two-point curvature correlator on the space $M$. 
Measurements of the normalized average sphere distances $\bar{d}_x/\delta$ and $\bar{d}_y/\delta$ for spheres $S_x^\delta$ and $S_y^\delta$ of the
same radius $\delta$ will be correlated, for $\delta\!\in\! [8,18]$.}
\label{fig:spherescorr}
\end{figure}

\subsection{Measuring curvature correlators}
\label{subsec:curv}

With all necessary ingredients in place, we can now investigate the normalized, diffeomorphism-invariant connected two-point curvature correlators of 2D CDT. 
In accordance with our discussion of the quantum Ricci curvature or, more precisely, the quantum Ricci scalar in Sec.\ \ref{sec:qrc}, 
the (quasi-)local scalar quantity associated with a vertex $x$ whose correlator we will consider is the normalized average sphere distance 
$\bar{d}(S_x^{\delta},S_{x}^{\delta})/\delta$ of eq.\ (\ref{discnasd}), for which we will also use the shorthand $\bar{d}_x(\delta)/\delta$ or just 
$\bar{d}_x/\delta$ for notational simplicity.

We have already seen that the expectation value of the
manifold average of this quantity has a specific, nontrivial $\delta$-dependence (Fig.\ \ref{fig:curvprofile}). 
To determine the corresponding two-point correlation functions involves measuring 
$\bar{d}/\delta$ at a given sphere radius $\delta$ for point pairs $(x,y)$, which are a geodesic link distance $r$ apart, 
as illustrated schematically by Fig.\ \ref{fig:spherescorr}. 
In the present study, we will average over all point pairs, and not distinguish between space- and time-like separations between points, 
to the extent these concepts can be said to be ``inherited" from the Lorentzian geometries before the Wick rotation, which is not completely
straightforward.

Recall that the computation of the average distance $\bar{d}(\delta)$ of a sphere $S^\delta$ to itself
depends on the patch of geometry contained inside the sphere. We therefore expect to find correlations between 
the quasi-local quantities $\bar{d}_x(\delta)$ and $\bar{d}_y(\delta)$ whenever the two spheres intersect or touch, i.e.\ whenever $r\!\leq\! 2\delta$,
and possibly slightly beyond.
Of course, these correlations are spurious in the sense that they are there by construction, due to the fact that we need a whole vertex neighbourhood
to determine the local curvature. They will not be considered for the continuum analysis.

\begin{figure}[t]
\centering
\includegraphics[width=0.8\textwidth]{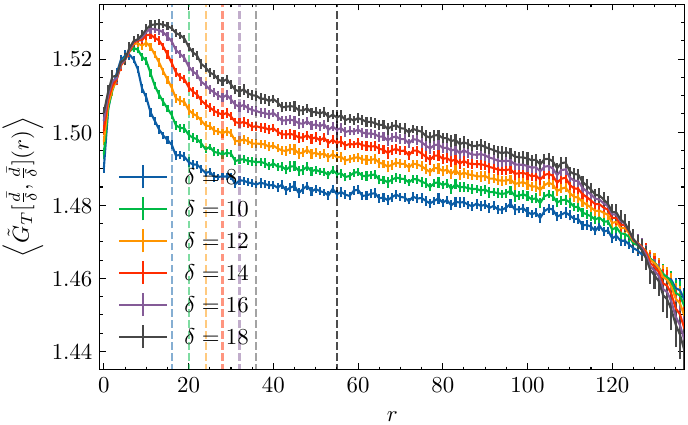}
\caption{Normalized curvature correlator $\langle \tilde{G}_T[\tfrac{\bar{d}}{\delta},\tfrac{\bar{d}}{\delta}](r)\rangle$ as a function of the link distance $r$, 
for volume $N_2\! =\! 300$k and various sphere radii $\delta\!\in\! [8,18]$. The vertical dashed line of matching colour indicates the value $r\! =\! 2\delta$ below which
we expect correlations to be spurious. The black vertical dashed line marks the radius $r_\mathrm{max}\! =\! 55$ below which finite-size effects are negligible.
}
\label{fig:corr1}
\end{figure}

All simulations described in this section were performed at fixed two-volumes $N_2\! =\! 100$k, 200k and 300k, 
and sphere sizes of even radius $\delta$ in the interval $\delta\!\in\! [8,18]$. 
This range corresponds to the initial rise of the curvature profile beyond the region of
lattice artefacts for small $\delta$. 
For each volume used, we have determined a maximal distance $r_\mathrm{max}$, below which finite-size effects are considered 
negligible, as explained further in Appendix A. The data points relevant 
for a physical continuum interpretation therefore lie in the region $2\delta\! <\! r\! \lesssim \! r_\mathrm{max}$.

The ensemble average was estimated by using 150 configurations sampled via the Markov chain Monte Carlo algorithm, 
where we performed $N_2\!\times\! 200$ Monte Carlo moves between subsequent measurements.
The statistical errors have been estimated using a batched bootstrapping method to account for possible correlations between successive samples of 
configurations in the Monte Carlo algorithm.
Since our moves are rejection-free, this means
that each site of the geometry was on average updated 200 times. In each of the 150 triangulations, the correlator
(\ref{corr1}) was estimated by sampling 5000 vertex pairs uniformly. To the best of our knowledge, our methodology for implementing a uniform sampling is new, 
and different from related prescriptions in the literature, which seem to contain a systematic bias. This is explained in more detail in Appendix B.

\begin{figure}[t]
\centering
\includegraphics[width=0.8\textwidth]{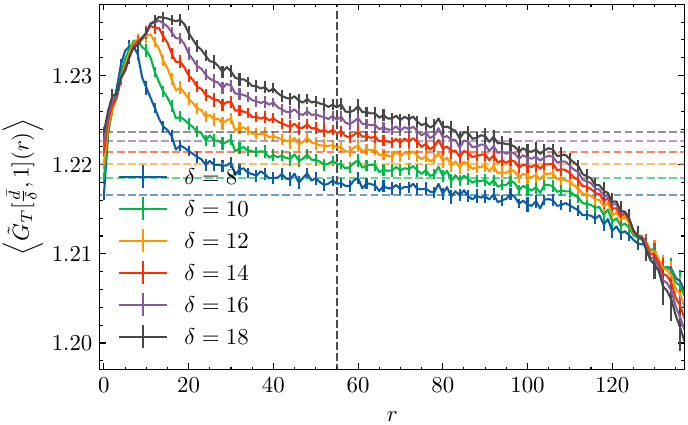}
\caption{Normalized correlator $\langle \tilde{G}_T[\tfrac{\bar{d}}{\delta},1](r)\rangle$ as a function of the link distance $r$, 
together with the ensemble average $\big\langle \overline{\bar{d}/\delta}|_T \big\rangle$ (dashed horizontal line of matching colour),
for volume $N_2\! =\! 300$k and various sphere radii $\delta\!\in\! [8,18]$.}
\label{fig:corr2}
\end{figure}

We start by reporting the results of the normalized curvature correlator 
\begin{equation}
\langle \tilde{G}_T[\tfrac{\bar{d}}{\delta},\tfrac{\bar{d}}{\delta}](r)\rangle =\bigg\langle \frac{G_T[\tfrac{\bar{d}}{\delta},\tfrac{\bar{d}}{\delta}](r)}{G_T[1,1](r)}
\bigg\rangle .
\label{corr1}
\end{equation}
The plot of the measurement data for the largest volume $N_2\! =\! 300$k is shown in Fig.\ \ref{fig:corr1}.
We observe that for each $\delta$, the measured correlator (\ref{corr1}) starts out with a peak that lies almost entirely inside 
$r\!\leq\! 2\delta$, then decreases gently as a function of $r$ throughout the trusted $r$-range, before taking a steeper drop around $r\!\approx\! 110$, 
far beyond $r_\mathrm{max}$.
Unsurprisingly, the width and height of the initial peak grows for larger values of $\delta$, indicating stronger correlations due to the larger size of the overlap
region between the interiors of the two spheres. 

Next, let us look at the measurements of the normalized correlator of the curvature with the unit operator,
\begin{equation}
\langle \tilde{G}_T[\tfrac{\bar{d}}{\delta}, 1 ](r)\rangle =\bigg\langle \frac{G_T[\tfrac{\bar{d}}{\delta}, 1 ](r)}{G_T[1,1](r)}
\bigg\rangle =\bigg\langle \frac{\sum_{x,y\in T} \tfrac{\bar{d}_x}{\delta}\, \delta_{d(x,y),r} }{ \sum_{x,y\in T}  \delta_{d(x,y),r}}\bigg\rangle,
\label{corr2}
\end{equation}
shown in Fig.\ \ref{fig:corr2}. As anticipated in Secs.\ \ref{subsec:class} and \ref{subsec:cdt}, 
due to the non-constancy of the geometry there are nontrivial correlations between the average sphere distance $\bar{d}_x$ and the
Kronecker delta $\delta_{d(x,y),r}$ in $G_T[\tfrac{\bar{d}}{\delta}, 1 ](r)$, such that (\ref{corr2}) is not equal to the
ensemble average $\big\langle \overline{\bar{d}/\delta}|_T\big\rangle$, which we have included in the figure for comparison. Note that
the shapes of the curves for $\langle \tilde{G}_T[\tfrac{\bar{d}}{\delta}, 1 ](r)\rangle$ in Fig.\ \ref{fig:corr2} are rather similar to those for
$\langle \tilde{G}_T[\tfrac{\bar{d}}{\delta},\tfrac{\bar{d}}{\delta}](r)\rangle$ in Fig.\ \ref{fig:corr1}. We will analyze next how they combine
when we determine the connected correlator. 

\begin{figure}[t]
\centering
\includegraphics[width=0.9\textwidth]{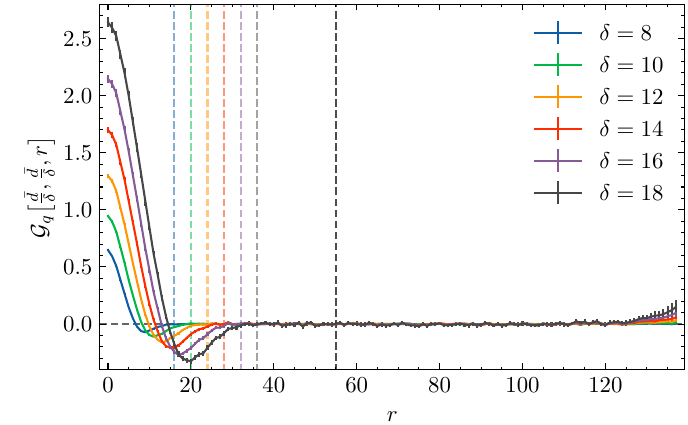}
\caption{Normalized connected curvature correlator ${\cal G}_q[\tfrac{\bar{d}}{\delta},\tfrac{\bar{d}}{\delta},r] $ as a function of the link distance $r$
for volume $N_2\! =\! 300$k and various sphere radii $\delta\!\in\! [8,18]$.}
\label{fig:corr3}
\end{figure}

Fig.\ \ref{fig:corr3} shows the Monte Carlo measurements for the connected curvature correlator
\begin{equation}
{\cal G}_q[\tfrac{\bar{d}}{\delta},\tfrac{\bar{d}}{\delta},r] = \langle \tilde{G}_T[\tfrac{\bar{d}}{\delta},\tfrac{\bar{d}}{\delta}](r)\rangle -
2  \big\langle \overline{\bar{d}/\delta}|_T\big\rangle \langle \tilde{G}_T[\tfrac{\bar{d}}{\delta}, 1 ](r)\rangle +\big\langle \overline{\bar{d}/\delta}|_T\big\rangle^2,
\label{corr3}
\end{equation}
whose general form
was defined in eq.\ (\ref{qopt1}). The most striking feature of this correlator is that it vanishes identically within measuring accuracy outside 
$r\!\leq 2\delta$, independent of the value of $\delta$ for which we measure the average sphere distance. In other words, we find a perfect
cancellation among the nonvanishing terms on the right-hand side of eq.\ (\ref{corr3}), which even extends to distances $r$ significantly above the trusted range
$2\delta\! <\! r\!\lesssim\! 55$. The deviations from zero that appear for $r\! \gtrsim\! 120$ are due to finite-size effects.\footnote{We have also measured the connected
correlators for smaller volumes $N_2$, where the results are similar, but the finite-size effects set in at smaller values of $r$.}
For small $r$, $0\!\leq\! r\!\lesssim\!2\delta$, we observe a qualitatively similar behaviour for all values of $\delta$,
namely an initially positive correlation, which for growing $r$ dips briefly into negative territory before going to zero
at $r\!\approx\! 2\delta$. We do not have an explanation for the detailed behaviour of these curves; correlations are clearly present, caused by the overlap
of the interiors of the spheres, but these are mixed with short-distance discretization artefacts, which are known to affect the average sphere distance
for small $\delta$. At any rate, we do not expect this region to be associated with interesting continuum physics.

Note that a ``na\"ive" subtraction
\begin{equation}
\langle \tilde{G}_T[\tfrac{\bar{d}}{\delta},\tfrac{\bar{d}}{\delta}](r)\rangle -\big\langle \overline{\bar{d}/\delta}|_T\big\rangle^2,
\label{naive}
\end{equation}
which would be appropriate for a connected correlator of local quantum fields on a fixed, flat background, in our case
does not produce a vanishing result, because the first term
has a nontrivial $r$-dependence (cf.\ Fig.\ \ref{fig:corr1}), while the second one is constant.
Since the correlator vanishes inside the relevant $r$-ranges for all volumes we have investigated, a finite-size scaling analysis, where one overlaps measurements after
rescaling distances according to $r\rightarrow r/N_2^{1/d_H}\! =\! r/N_2^{1/2}$, is straightforward and compatible with a vanishing continuum correlator. 

From a physical viewpoint the vanishing of the curvature correlator ${\cal G}_q[\tfrac{\bar{d}}{\delta},\tfrac{\bar{d}}{\delta},r]$ 
has a clear-cut interpretation: because of the absence of local propagating degrees 
of freedom in two-dimensional quantum gravity, we do not expect to find correlations between \textit{any} local geometric operators. Our measurements
confirm this expectation and at the same time strengthen confidence in the construction of manifestly diffeomorphism-invariant correlators in
nonperturbative quantum gravity, and the feasibility of computing correlators of the quantum Ricci curvature in particular.

\section{Summary and outlook}
\label{sec:future}

We have presented above the first quantitative analysis of diffeomorphism-invar\-iant 
two-point correlators of the curvature in nonperturbative Lorentzian quantum gravity, 
albeit in a two-dimensional toy model.
Another novel feature was the use in the correlator of the quantum Ricci curvature, a notion of curvature geared towards nonperturbative lattice quantum gravity, with an 
improved UV behaviour compared to other prescriptions \cite{curvsumm}. We have demonstrated that the construction and numerical implementation of these
correlators is conceptually and technically feasible and leads to a physically sensible result, namely the vanishing of the connected curvature correlator (\ref{corr3}) 
within measuring accuracy. 

Our analysis highlighted unexpected structural features of the connected correlators, which are due to the dynamical nature of the geometry and
would not be present on a fixed, flat background space(-time). They lead to ambiguities in the definition of these correlators, which turned out to be immaterial
in 2D Lorentzian quantum gravity, but may well be important in other cases. 

With the study of correlators in nonperturbative quantum gravity still in its infancy, the successful completion of our two-dimensional implementation 
is very encouraging for their further development. 
As mentioned in the introduction, our main motivation is the investigation of curvature correlators in full, four-dimensional quantum gravity in a near-Planckian
regime, using the technical framework of lattice gravity based on CDT. 
From a cosmological perspective, it would be very interesting to measure 
correlators associated with spacelike hypersurfaces in the emergent de Sitter-like quantum universe found in this formulation. 
In standard early-universe cosmology, such correlators describe the quantum fluctuations of a hypothetical inflaton field, which act as seeds of primordial 
inhomogeneities in a de Sitter-like background spacetime, and are also directly related to the so-called power spectrum \cite{mukhanov,relcos}. 
If we can derive correlators and their properties from first principles in nonperturbative quantum gravity, this may provide a 
purely quantum-gravitational explanation and possibly new predictions for the nature of these fluctuations in cosmology,    
without resorting to additional matter fields. This is the subject of ongoing research we hope to report on elsewhere.

\vspace{0.2cm}

\subsection*{Acknowledgments} 
RL thanks the Perimeter Institute for hospitality.
This research was supported in part by Perimeter Institute for Theoretical Physics. Research at Perimeter Institute is supported by the 
Government of Canada through the Department of Innovation, Science and Economic Development and by the Province of Ontario through
the Ministry of Colleges and Universities.

\section*{Appendix A}
In this appendix, we explain the method we have used to establish a range of distances $r\!\lesssim\! r_\mathrm{max}$ 
where finite-size effects in the measurements of the correlators $\langle \tilde{G}(r)\rangle$, ${\cal G}_q[r]$
presented in Sec.\ \ref{subsec:curv} are negligible. 
It is related to what we call the \textit{effective linear size} $L$ of a typical triangulation in an ensemble at fixed two-volume. 
This is an approximate concept, which we obtain by monitoring the discrete volume of a 2D ball $B^r_x$ (all triangles contained inside the sphere $S^r_x$),
as one would do when extracting the Hausdorff dimension $d_H$ from the leading scaling behaviour of the expectation value
$\langle \overline{\mathrm{vol}(B^r)}\rangle\! \propto\! r^{d_H}$. As we increase the ball radius $r$ on triangulations of a given volume $N_2$, 
this scaling will start deviating from a power law once the typical ball wraps around the spatial direction of the torus and starts to selfoverlap.\footnote{The
time extensions $t_\mathrm{tot}$ of our triangulations are such that wrapping in the spatial direction always happens much before it would occur in
the time direction, see also footnote \ref{fn4}.} 
We then identify the length $L$ as the diameter $2r_\mathrm{max}$ of the largest typical ball whose selfoverlaps are sufficiently small not to
affect the (quadratic) power-law scaling of its volume significantly. 
The linear size $L$ is therefore also a rough estimate of the length of a typical shortest, noncontractible geodesic that wraps once around the spatial direction of the torus. 
We have made rather conservative choices, such that $r_\mathrm{max}$ is located well below the radius
where a deviation from a single power law becomes visible in the volume plots (see \cite{vdduin} for technical details). 
Half of the linear scale $L$ then gives us a maximal distance $r_\mathrm{max}=L/2$ below which we are confident that the influence of finite-size
effects on the correlators is negligible. For the volumes $N_2\! =\! \{50\mathrm{k},100\mathrm{k},150\mathrm{k},200\mathrm{k},250\mathrm{k},300\mathrm{k}\}$, 
the maximal distances obtained in this way are
$r_\mathrm{max}\! =\! \{31,44,49,51,53,55\}$ respectively.

\section*{Appendix B}\label{pps}
An important element of the computational implementation of the two-point correlators described in the main text is a uniform sampling of point pairs 
$(x,y)$ at a given discrete distance $r$.
It is feasible to evaluate the average over all point pairs $(x, y)$ fully, but it is significantly more effective computationally to estimate this average with a sample of point pairs.
Uniformity of this sampling is defined with respect to the volume weight, which is 1 for each vertex. 
Accordingly, all pairs of vertices $(x,y)$, $d(x,y)\! =\! r\!\not=\! 0$, should also have equal weight and be selected with the same probability. Since the underlying geometries
are irregular random triangulations, such a uniform sampling is slightly nontrivial.
A straightforward implementation for a given geometry $T$ 
would be to create a complete list of all vertex pairs at distance $r$, and sample uniformly from this set. 
However, creating such a list requires identifying all vertex pairs, in which case one could perform the full sum over all point pairs without additional computational cost.
In this case, one would not gain any computational benefit from sampling, especially since we tend to require only relatively small samples to estimate manifold averages.

Note that the same sampling issue arises in measurements of the (average) quantum Ricci curvature, whenever one uses the prescription involving
the average sphere distance of two $\delta$-spheres whose centres are a distance $\delta$ apart, obtained by setting $\varepsilon\! =\! \delta$ in
the defining relation (\ref{sdist}). In this situation, one also needs to sample vertex pairs $(p,p')$ at a prescribed distance. 
This was implemented in
previous work \cite{qrc2,qrc3,Brunekreef2021} by (i) uniformly sampling a first vertex $p\!\in\! T$, (ii) constructing the $\delta$-sphere $S^\delta_p$
around $p$, and (iii) selecting a second vertex $p'$ uniformly from $S^\delta_p$.
This is a much more efficient method, since it only requires a single
breadth-first search for every vertex pair (as opposed to $N_0(T)$ breadth-first searches for all vertex pairs), but it does not amount to a uniform
sampling of point pairs! Instead, the second vertex $p'$ is chosen with a relative weight $1/\mathrm{vol}(S^\delta_p)$, which depends on $p$. These
weights would all be identical if the geometry was completely regular and the neighbourhood of each vertex looked the same, but this is of course generally not the case. 

To what extent the bias resulting from this nonuniform sampling is significant, in the sense of affecting continuum outcomes, is not immediately clear.
In general, its effect will depend both on the observable studied and the nature of the typical geometry in the ensemble. 
Previous studies of correlators in EDT \cite{bakkersmit,bialas} do not specify their methodology explicitly, but may also have used a nonuniform sampling.   
With growing vertex distance,
the bias can decrease if the individual sphere volumes become more similar to their manifold average. We have found some evidence for the latter in
the 2D CDT data, but it is not sufficiently stringent to draw any general conclusion about the equivalence or otherwise of the two sampling methods.
We have not attempted a comprehensive analysis of their difference either, but have repeated the measurement of the
curvature profile $\langle \bar{d}_\mathrm{av}(\delta)/\delta \rangle$ for $\varepsilon \! =\! \delta$ in 2D CDT of reference \cite{Brunekreef2021} 
with a uniform sampling. The main net effect outside the region of lattice artefacts is a small vertical upward shift of the data points, compared to the
nonuniform sampling \cite{vdduin}. 
Since the vertical offset of the curvature profile is known to be not universal, the influence of the samplings on this particular observable 
seems rather mild, at least in the range $N_2\!\leq\! 300$k of volumes considered, which is reassuring with regard to the validity of the earlier
investigations of curvature profiles. 

Having pinpointed the origin of the nonuniformity, we can modify the corresponding sampling in a straightforward way to make it ``effectively
uniform" while still taking advantage of its computational simplicity, by including compensating weight factors. The modified sampling prescription is as follows:
(i)$'$ pick a first vertex $x\!\in\! T$ uniformly at random, (ii)$'$ construct the $r$-sphere $S^r_x$ around $x$, (iii)$'$ select a second vertex $y$ uniformly 
from $S^r_x$, and (iv)$'$ multiply the measurement associated with the point pair $(x,y)$ by a weight factor $\mathrm{vol}(S^r_x)/\overline{\mathrm{vol}(S^r)}|_T$.
There is no extra computational cost in this last step, since the sphere $S^r_x$ has to be determined in any case. We have tested this weighted sampling
method for the two-point correlator of the coordination number $c(x)$ in 2D CDT, and verified that within statistical error bars it leads to identical results for
the same quantity sampled uniformly by using complete lists of point pairs, as described earlier. All correlator measurements in this paper have been
performed by using the weighted, effectively uniform sampling.

\end{document}